\documentclass[12pt]{article}
\usepackage[mathscr]{eucal} 
\usepackage{amscd}
\usepackage{amssymb,latexsym}
\usepackage{verbatim}
\usepackage{graphicx}

\newtheorem{thm}{Theorem}[section]
\newtheorem{Def}{Definition}[section]
\newtheorem{prop}[thm]{Proposition}
\newtheorem{cor}[thm]{Corollary}
\newcommand{\qed}{\hfill $\Box$ \medskip}
\newcommand{\proof}{\noindent \emph{Proof:\ }}

\renewcommand\S{\Sigma}
\newcommand\s{\sigma}
\renewcommand\d{\partial}

\newcommand\D{\nabla}
\newcommand\e{\epsilon}

\renewcommand\r{\rho}

\newcommand\g{\gamma}

\newcommand\scri{\mathscr{I}}

\newcommand\beq{\begin{eqnarray}}
\newcommand\eeq{\end{eqnarray}}
\newcommand\ben{\begin{enumerate}}
\newcommand\een{\end{enumerate}}
\newcommand\bit{\begin{itemize}}
\newcommand\eit{\end{itemize}}

\newcounter{mnotecount}[section]

\title{Cosmological Spacetimes with $\Lambda >0$}

\author{Gregory J. Galloway\thanks{Supported in part by NSF grant \# DMS-0104042} \\
 Department of Mathematics\\University of Miami}
\begin{document}
\date{}
\maketitle

\section{Introduction}

This paper is based on a talk given by the author at the BEEMFEST, held at the University of Missouri, Columbia,
in May, 2003.  It was indeed a great pleasure and priviledge to participate in this meeting
honoring John Beem, whose many oustanding contributions to mathematics, and leadership 
in the field of Lorentzian geometry has enriched so many of us.

In this paper we present some results, based on joint work with Lars Andersson~\cite{AG},  
concerning certain
global properties of asymptotically de Sitter spacetimes satisfying the Einstein equations
with positive cosmological  constant.  
There has been increased interest in such spacetimes in 
recent years due, firstly,  to observations concerning the rate of expansion of the universe, suggesting the 
presence of a positive cosmological constant in our universe,
and, secondly, due to recent efforts to understand quantum gravity on de
Sitter space via, for example, some de Sitter space version of the AdS/CFT correspondence; cf., \cite{bou} 
and references cited therein.  In fact, the results discussed here were originally motivated by results
of Witten and Yau \cite{WY,CG}, and related spacetime results on topological censorship \cite{GSWW},
pertaining to the AdS/CFT correspondence.  These  papers give illustrations of how the geometry and/or
topology of the conformal infinity of an asymptotically hyperbolic Riemannian manifold, or
an asymptotically anti-de Sitter spacetime, can influence the global structure of the manifold.
In this vein, the results described here establish connections between  the geometry/topology of
conformal infinity and the occurence of singularities in asymptotically de Sitter spacetimes.

\section{Asymptotically de Sitter spacetimes}

We use Penrose's notion of conformal infinity \cite{Psimple} to make precise the notion of
``asymptotically de Sitter". Recall, this notion is based on the way in which the standard Lorentzian space forms, Minkowski
space, de Sitter space and anti-de Sitter space, conformally embed into the Einstein static universe $(\Bbb R
\times S^n, -du^2 + d\omega^2)$.  For concreteness, throughout the paper we restrict attention to spacetimes
$(M^{n+1},g)$ (of signature $(-+\cdots +)$), which are globally hyperbolic, with {\it compact} Cauchy surfaces.
We shall refer to such spacetimes as {\bf cosmological spacetimes}.  We refer the reader to the 
core references \cite{BE,HE,ON}
for basic results, terminology and notation in Lorentzian geometry and causal theory.

Recall, de Sitter space is the unique geodesically complete, simply connected spacetime of constant
{\it positive} curvature ($=1$ in equation (\ref{deS}) below). As such,
it is a vacuum solution to the Einstein equations with positive cosmological constant (see the next section
for further discussion of the Einstein equations).  It
can be represented as a hyperboloid of one-sheet in Minkowski space, and may be expressed in global coordinates as
the warped product,
\beq\label{deS}
M = \Bbb R \times S^n, \qquad ds^2 = -dt^2 + \cosh^2t \,d\omega^2  \,.
\eeq
 Under the transformation 
$u = \tan^{-1}(e^t) - \pi/4$, the metric (\ref{deS}) becomes
\beq
ds^2  = \frac1{\cos^2(2u)} (-du^2 +d\omega^2)  \,.
\eeq   
Thus, de Sitter space conformally embeds onto the region $-\pi/4 < u < \pi/4$ in the Einstein static
universe;
see Figure 1. 
Future conformal infinity $\scri^+$ (resp., past conformal infinity $\scri^-$) is represented
by the \emph{spacelike} slice $u = \pi/4$ (resp., $u= -\pi/4$).  This situation
serves to motivate the following definitions.

\vspace{.1in}
\begin{center}
\mbox{
\includegraphics[width=1.4in]{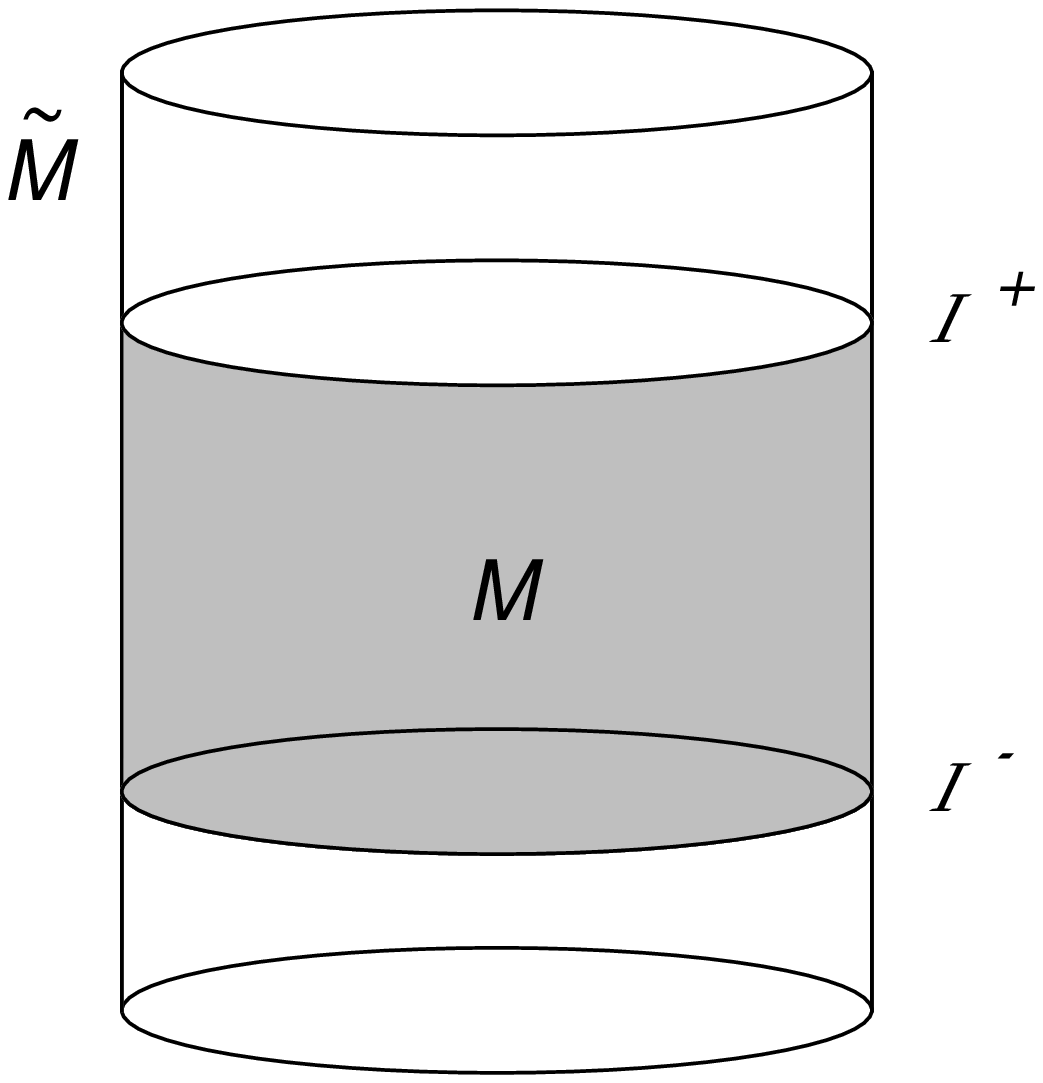}
}
\\ \begin{small}
\vspace{.1in}\textsc{Figure 1.} Conformal embedding of de Sitter space. \end{small}
\end{center}

\begin{Def}\label{asymp}
A cosmological spacetime $(M,g)$  is said to be {\bf future asymptotically de Sitter}  provided
there exists a spacetime-with-boundary
$(\tilde M,\tilde g)$  and a smooth function $\Omega$ on $\tilde M$ such that
(a) $M$ is the manifold interior of $\tilde M$, (b) the manifold boundary $\scri^+ = \d \tilde
M$ ($\ne \emptyset$) of $\tilde M$  is spacelike and lies to the future of $M$, $\scri^+ \subset I^+(M,\tilde
M)$, and (c) 
$\tilde g$ is conformal to 
$g$, i.e., $\tilde g = \Omega^2 g$, where $\Omega = 0$ and $d\Omega \ne 0$ along $\scri^+$.
\end{Def}
We refer to $\scri^+$ as future (conformal) infinity.
In general, no assumption is made about the topology of $\scri^+$; in particular it need not be compact.  
The conformal factor $\Omega$ is also referred to as the {\it defining function} for $\scri^+$.

\begin{Def}
A future asymptotically de Sitter spacetime $M$ is said to be  {\bf future asymptotically simple} 
provided every future
inextendible null geodesic in $M$ reaches future infinity, i.e. has an end point on $\scri^+$.
\end{Def}

It is a basic fact \cite{AG} that a cosmological spacetime $M$ is future asymptotically
simple if and only if $\scri^+$ is compact; in this case $\scri^+$ is diffeomorphic to the Cauchy
surfaces of $M$.  {\bf Past asymptotically de Sitter} and {\bf past asymptotically simple} are defined in a
time-dual manner.  $M$ is said to be asymptotically de Sitter provided it is both past and future
asymptotically de Sitter.

\smallskip
{\it Schwarzschild-de Sitter space.} Schwarzschild-de Sitter (SdS) space  is a good
example of  a spacetime that is asymptotically de Sitter but not asymptotically simple.  Roughly speaking 
it represents a Schwarzschild black hole sitting in a de Sitter background.  Its metric in 
static coordinates (and in four dimensions) is given by,
\beq
\qquad 
ds^2 = -(1 -\frac{2m}{r} -\frac{\Lambda}3 r^2)dt^2 + \frac1{(1 -\frac{2m}{r} -\frac{\Lambda}3 
r^2)}dr^2+r^2d\omega^2
\eeq 
where $\Lambda >0$, and $d\omega^2$ is the standard metric on the unit $2$-sphere.  
SdS space has Cauchy surface topology $S^1\times S^2$, while $\scri^+$ for
this spacetime has topology $\Bbb R\times S^2$. This spacetime fails to be asymptotically simple
(either to the past or future), since  there are null geodesics which enter the black hole region
to the future, and the white hole region to the past.
The Penrose diagram for this spacetime is given in Figure 2.

\begin{center}
\mbox{
\includegraphics[width=2.3in]{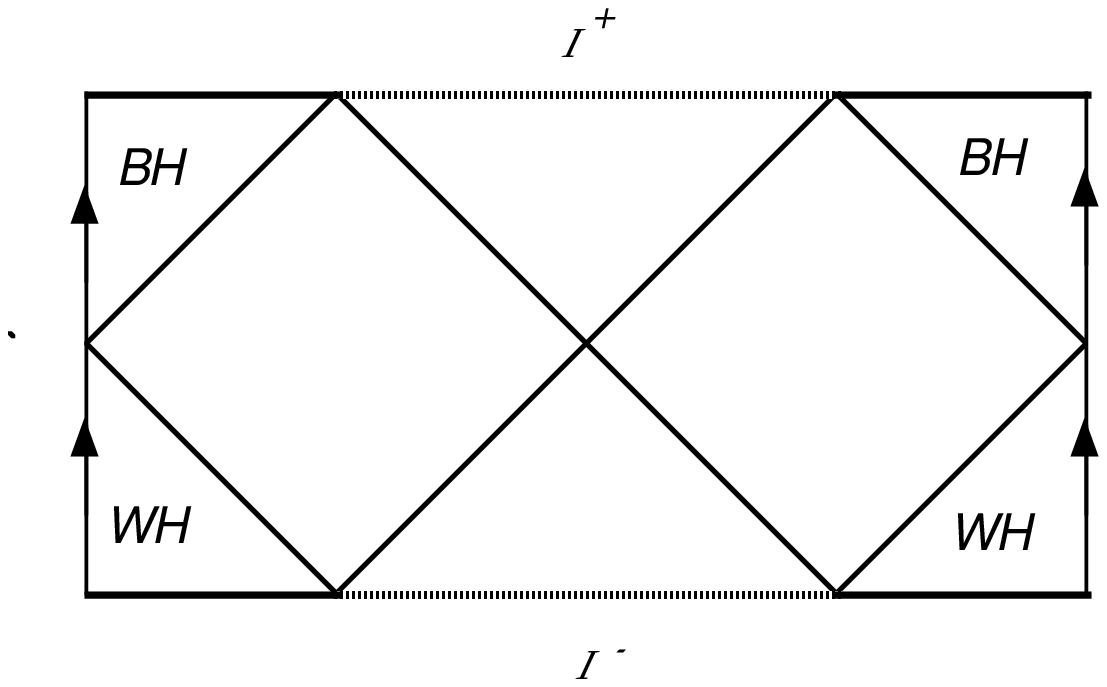}
}
\vspace{.1in}
\\ {\small \textsc{Figure 2.} Penrose diagram for Schwarzschild-de Sitter space. 
\\ \hspace*{-.25in}}
\end{center}

{\it The de Sitter cusp}.  By the de Sitter cusp, we mean the spacetime, given by,
\beq
M = \Bbb R \times T^n\,, \qquad ds^2 = -dt^2 + e^{2t} d\s_0^2
\eeq 
where $d\s^2_0$ is a flat metric on the $n$-torus $T^n$.  This spacetime is obtained as a quotient
of a region in de Sitter space.  In fact, the universal cover of this spacetime is isometric
to the ``half" of de Sitter space shown in Figure 3.    
With the exception of the $t$-lines, all timelike geodesics in the de Sitter cusp are past incomplete.
This spacetime is future asymptotically de Sitter (and future asymptotically simple, as well), but 
is not past asymptotically de Sitter.  

\begin{center}
\mbox{
\includegraphics[width=1.75in]{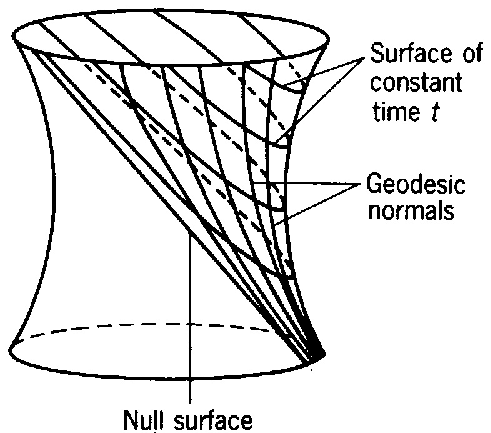}
}\\
\vspace{.07in}{\small \textsc{Figure 3.} The universal cover of the de Sitter cusp.}
\end{center}  
\vspace{.1in}

The examples considered above are vacuum solutions to the Einstein equations with positive
cosmological constant.  In the next section we consider some matter filled models.

\section{Occurrence of singularities and the geometry of conformal infinity}

In this section we consider spacetimes $M^{n+1}$ which obey the 
Einstein equations,
\beq\label{einstein}
R_{ij} -\frac12Rg_{ij} +\Lambda g_{ij} = 8\pi T_{ij} \,,
\eeq
with {\it positive} cosmological constant $\Lambda$, where the energy-momentum tensor
$T_{ij}$  will be assumed to satisfy certain energy inqualities.  

Following (more or less) conventions in \cite{HE}, $M$ is said to obey 
\ben
\item[(a)] the {\it strong energy condition} provided, 
\beq\label{strong}
(T_{ij} - \frac1{n-1}Tg_{ij})X^iX^j \ge 0
\eeq
for all timelike vectors $X$, where $T = T_i{}^i$,
\item[(b)] the {\it weak energy condition condition} provided,
\beq\label{weak}
T_{ij}X^iX^j \ge 0
\eeq 
for all timelike vectors $X$, 
\item[(c)] the {\it dominant energy condition} provided,
$T_{ij}X^iY^j \ge 0$
for all causal vectors $X$ and $Y$ that are either both future pointing or both past pointing, and
\item[(d)] the {\it null energy condition} provided (\ref{weak}) holds for all {\it null} vectors $X$.
\een

\smallskip
Setting $\Lambda = n(n-1)/2{\ell^2}$, the strong energy condition~(\ref{strong}) is
equivalent to,
\beq\label{ricstrong}
{\rm Ric}\,(X,X) = R_{ij}X^iX^j \ge -\frac{n}{\ell^2}
\eeq 
for all unit timelike vectors $X$. 
Finally, note, the null energy condition  is equivalent to,
\beq\label{ricnull}
{\rm Ric}\,(X,X) = R_{ij}X^iX^j \ge 0
\eeq
for all null vectors $X$.

We now consider  some classical dust filled FRW models which are solutions to the Einstein equations
(\ref{einstein}); see, e.g., \cite[chapter 23]{dinverno}.  Thus, let $M$ be a warped product spacetime of
the form,
\beq\label{frw}
M= \Bbb R\times \S^3, \qquad ds^2 = -dt^2 + R^2(t)d\s_k^2 
\eeq
where $(\S^3,d\s_k^2)$ is a compact Riemannian manifold of constant curvature $k=-1, 0, +1$.  
Let the 
energy-momentum tensor of $M$ be that corresponding to a dust, i.e., pressureless perfect fluid,
\beq
T_{ij} = \r\, u_iu_j\, ,
\eeq
where $\r= \r(t)$ and $u = \d/\d t$.  The Einstein equations imply that the scale factor
$R=R(t)$ obeys the Friedmann equation,
\beq
(R')^2 = \frac{C}{R} + \frac13\Lambda\, R^2 - k  \,,
\eeq
where,
\beq\label{mass}
\frac83\pi R^3\rho = C =\mbox{ constant }  \, .
\eeq

The table in Figure 4 summarizes the qualitative behavior of the scale factor $R(t)$ 
for all choices of the sign of the cosmological constant $\Lambda$ and of the sign of $k$. Focussing 
attention on the column $\Lambda > 0$,  and those solutions for which the scale factor is unbounded to the future, 
we observe that
the positively curved case
$k=+1$ is distinguished from the  nonpositively curved cases $k= -1,0$. 
In the  case $k=+1$, if the ``mass parameter" $C$ (which we assume to be positive)
is sufficiently small, relative to the cosmological constant $\Lambda$, then 
$M$ is timelike geodesically  complete to the past,
as well as the future. (In Figure 4, $\Lambda_c =4/9C^2$.)  In this regard, $M$ behaves like de Sitter space, which in fact
corresponds to the limiting case
$C=0$.  On the other hand, in the cases $k=-1,0$, no matter how small the mass parameter $C$, $M$ begins
with a big bang singularity, and hence all timelike geodesics in $M$ are past incomplete.

\vspace{.05in}
\begin{center}\hspace*{-.15in}
\mbox{
\includegraphics[width=5.1in]{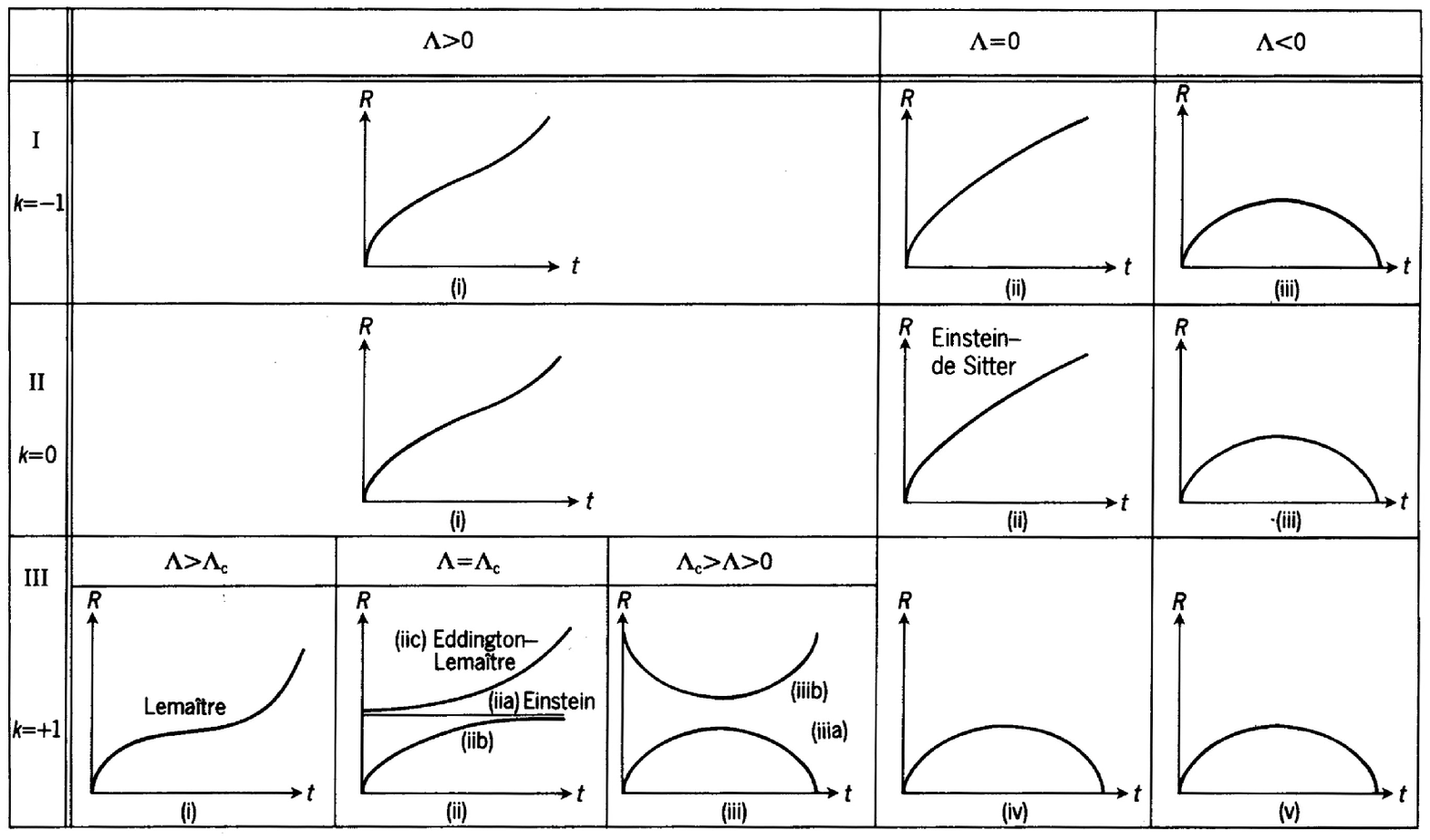}
}\\
\vspace{.1in} {\small \textsc{Figure 4.} Behavior of the scale factor $R(t)$}.\footnote{We thank Ray D'Inverno
\cite{dinverno} and Oxford University Press for permission to reproduce this chart.} 
\end{center}  
\vspace{.1in}

The aim of our first results is to show that the past incompleteness
in the cases $k=-1,0$ holds in a much broader context, one that does not require exact symmetries, 
assumptions of constant curvature, etc.  In all (unbounded) cases the scale  factor behaves like,
$R(t)\sim e^{\sqrt{\frac{\Lambda}3}t}$ for large $t$, just as in de Sitter space.  In fact, and this
is the key to the generalization, all of these models are future asymptotically simple and de Sitter,
in the sense of section 2.

Let $(M^{n+1},g)$ be a cosmological spacetime which is future asymptotically simple and de Sitter,
in the sense  of section 2.   Then there exists a spacetime with boundary $(\tilde M,\tilde g)$
such that $\tilde M = M \cup \scri^+$, where $\scri^+$ is compact (and connected), and  $\tilde g = \Omega^2
\,g$, as in Definition
\ref{asymp}. 
The Riemannian metric
$\tilde h$ induced by $\tilde g$ on $\scri^+$ changes by a
conformal factor
with a change in  the defining function $\Omega$, and thus
$\scri^+$ is endowed with a natural conformal structure
$[\tilde h]$.  By  the positive resolution of the Yamabe conjecture \cite{Schoen}, 
the conformal class $[\tilde h]$ contains
a metric of constant scalar curvature $-1$,
$0$, or $+1$, exclusively, in which case we will say that $\scri^+$
is of negative, zero, or positive Yamabe
class, respectively.  In this way, every future asymptotically simple  and de Sitter 
spacetime falls into precisely one of three classes.   
We note that future conformal infinity 
$\scri^+$ for the dust-filled FRW model (\ref{frw}) - (\ref{mass})  
is of negative, zero, or positive Yamabe class precisely when the metric 
$d\s^2_k$ has curvature $k = -1, 0$ or $+1$, respectively.   

We now present a singularity theorem for future asymptotically simple
and de Sitter spacetimes of negative Yamabe class (which is a slight variation of 
Theorem~3.2 in \cite{AG}). 

\begin{thm}\label{sing} Let $(M^{n+1},g)$ be a future asymptotically simple and de Sitter spacetime
satisfying the Einstein equations with $\Lambda > 0$, such that
\ben
\item[(i)] the strong and weak energy conditions hold, and
\item[(ii)] the fall-off condition, $T_i{}^i \to 0$ on approach to $\scri^+$, holds.
\een
Then, if $\scri^+$ is of negative Yamabe class, every timelike geodesic in $(M,g)$ is
past incomplete.
\end{thm}

\proof Theorem \ref{sing} may be viewed as a Lorentzian analogue of the main result
of Witten-Yau \cite{WY}; its proof is essentially a  ``Lorentzification" of the proof of the Witten-Yau
result given in \cite{CG}.  

The Einstein equations and the assumptions on the energy-momentum tensor imply that (\ref{ricstrong})
holds and  that $R \to n(n+1)/{\ell^2}$ on approach to $\scri$, where $R = R_i{}^i$ is the 
spacetime scalar curvature.  Thus, by a constant rescaling of the metric, we may assume 
without loss of generality that $(M,g)$ satisfies, 
\beq\label{ricn}
{\rm Ric}\,(X,X) \ge -n \,,
\eeq
for all unit timelike vectors $X$, and 
\beq\label{scalar}
R \to n(n-1) \quad \mbox{on approach to } \scri^+ \,.
\eeq

The first step of the proof involves some gauge fixing.  
Fix a metric $\g_0$ in the conformal class $[\tilde h]$ of $\scri^+$ of constant
scalar curvature $-1$.
By the formula relating the scalar curvature functions
of conformally related metrics,  
(\ref{scalar})  implies that $\tilde g(\tilde\D\Omega,\tilde\D\Omega)|_{\scri^+} = -1$.
Then, as is shown in \cite{AG} (see also \cite{Graham} for the similar Riemannian case),
the conformal factor $\Omega$ can be chosen so that this equality holds in a neighborhood of $\scri^+$. 
To put it in a slightly different way,  $\Omega =t$ can  be chosen so that in a
$\tilde g$-normal neighborhood 
$[0,\e) \times\scri^+$ of $\scri^+$, the physical metric $g$ takes the form
\beq\label{gauge}
g =   \frac1{t^2}\left(-dt^2 +\tilde  h_t\right ) \, ,
\eeq 
where $\tilde h_t$, $t\in [0,\e)$, is the metric on the slice $\S_t = \{t\} \times \scri^+$ induced by 
$\tilde g$, such that, in addition,  $\tilde h_0 = \g_0$. (If $\Omega$ is the given defining function, one
seeks  a new defining function of the form $e^u\Omega$ with the requisite properties; this leads to a first
order non-charactersitic PDE for $u$, with appropriate boundary condition, which can always be solved; see
\cite{AG, Graham} for details.)

Let $H = H_t$, $0< t <\e$, denote the mean curvature of $\S_t$ in $(M,g)$.  By our conventions,
$H = {\rm div}\,u$, where $u= -t\frac{\d}{\d t}$ is the future pointing unit normal to the
$\S_t$'s. 
$H= H_t$ and $\tilde H = \tilde H_t$ (the mean curvature of  $\S_t$ in $(M,\tilde g|M)$) are related by,
\beq\label{meancurv}
H = t \tilde H + n \,.
\eeq
In particular, $H|_{\S_t} > 0$ for $t$ sufficiently small.

The Gauss equation (in the physical
metric $g$) applied to each
$\S_t$, together with the Einstein equations, yields the constraint,
\beq\label{gauss}
H^2 &=& 2{\mathscr T}(u,u) +2\Lambda +|K|^2 - r \nonumber \\ 
&=& 2t^2{\mathscr T}(\frac{\d}{\d t},\frac{\d}{\d t}) + n(n-1) +|K|^2 - t^2 \tilde r \,,
\eeq
where $r = r_t$ (resp., $\tilde r = \tilde r_t$) is the scalar curvature of $\S_t$ in the
metric induced from $g$ (resp., $\tilde g$),  $K= K_t$ is the second fundamental form of 
$\S_t$ in $(M,g)$, and ${\mathscr T} = T_{ij}$ is the energy-momentum tensor.
 Since, by the Schwarz inequality,  $|K|^2 \ge ({\rm tr}\,
K)^2/n =H^2/n$, equation (\ref{gauss}) implies,
\beq\label{Hineq}
H^2 & \ge & n^2 + \frac{n}{n-1}t^2\left({\mathscr T}(\frac{\d}{\d t},\frac{\d}{\d t}) -\tilde r\right) \nonumber\\
& \ge & n^2 - \frac{n}{n-1}\, t^2\,\tilde r_t  \,,
\eeq
where in the second inequality we have used the weak energy condition (\ref{weak}). Thus, since $\tilde r_0=$ the
scalar  curvature of $(\scri^+,\tilde h_0=\g_0)=-1$,
(\ref{Hineq}) implies that $H|_{\S_t} > n$ for all $t>0$ sufficiently small.  This is the essential consequence
of the assumption that $\scri^+$ is of negative Yamabe class.  Theorem \ref{sing} is now
an immediate consequence of the following result.
\qed

\begin{prop}\label{hawk}
Let $M^{n+1}$ be a spacetime satisfying the energy condition,
$$
{\rm Ric}\,(X,X)  \ge -n
$$
for all unit timelike vectors $X$.  Suppose that $M$ has a smooth
compact Cauchy surface $\S$ with
mean curvature $H$ satisfying $H> n$.  Then every timelike geodesic in $M$
is past incomplete.
\end{prop}

\proof This result a straightforward extension of an old singularity theorem
of Hawking; see e.g., \cite[Theorem 55A, p. 431]{ON}.  Its proof makes use
of basic comparison theory. 

Fix $\delta > 0$ so that the mean curvature of $\S$ satisfies $H\ge
n(1+\delta)$.  Let
$\rho : I^-(\S) \to \Bbb R$ be the Lorentzian distance function to $\S$,
\beq
\rho(x) = d(x,\S) = \sup_{y\in \S}d(x,y)  \,;
\eeq
$\rho$ is continuous, and smooth outside the past focal cut locus of $\S$.
We will show that $\rho$ is bounded from
above,
\beq
\rho(x) \le \coth^{-1}(1+\delta) \qquad \mbox{for all } x\in I^-(\S)  \,.
\eeq
This implies that every past inextendible timelike curve with future end
point on $\S$ has
length $\le \coth^{-1}(1+\delta)$.

Suppose to the contrary, there is a point $q\in I^-(\S)$ such that
$d(q,\S)=\ell> \coth^{-1}(1+\delta)$.
Let $\gamma:[0,\ell]\to M$, $t\to\gamma(t)$, be a past directed unit speed
timelike geodesic from
$p\in \S$ to $q$ that realizes the distance from $q$ to $\S$. $\gamma$ meets
$\S$ orthogonally, and
because it maximizes distance to $\S$, $\rho$ is smooth on an open set $U$
containing
$\gamma\setminus \{q\}$.  

For $0\le t< \ell$, the slice $\rho = t$ is
smooth near the point
$\gamma(t)$; let $H(t)$ be the mean curvature, with respect to the
future pointing normal $\D\rho$, at $\gamma(t)$ of the slice $\rho
= t$.
$H=H(t)$ obeys the traced Riccati (Raychaudhuri) equation,
\beq\label{eq:ray}
H' = {\rm Ric}(\gamma',\gamma') +|K|^2\,,
\eeq
where $'=d/dt$, and $K$ is  the second fundamental form of $\S$.

Equation (\ref{eq:ray}), together with the inequalities  $|K|^2 \ge  H^2/n$,
${\rm Ric}(\gamma',\gamma')\ge -n$ and $H(0) \ge n(1+\delta)$, implies that
$\mathscr H(t):=H(t)/n$
satisfies,
\beq
\mathscr H' \ge \mathscr H^2 - 1, \qquad \mathscr H(0) \ge 1+\delta\,.
\eeq
By an elementary comparison with the unique solution to: $h' = h^2 -1$,
$h(0) = 1+\delta$, we
obtain $\mathscr H(t)\ge \coth(a-t)$, where $a = \coth^{-1}(1+\delta) <
\ell$, which implies
that $\mathscr H = \mathscr H(t)$ is unbounded on $[0,a)$, contradicting
the fact that $\mathscr H$ is
smooth on $[0,\ell)$. \qed

We now briefly consider the ``borderline" case in which $\scri^+$ is of zero Yamabe class.
As the example of the de Sitter cusp described in Section 2 shows, the analogue of
Theorem \ref{sing} can fail in this case, in that {\it some} timelike geodesics
may be past complete.  But, as the following
theorem shows, it can fail only under very special circumstances.

\begin{thm}\label{sing2} Let $(M^{n+1},g)$ be a maximal future asymptotically simple and de Sitter spacetime
satisfying the Einstein equations with $\Lambda =n(n-1)/2$, such that
\ben
\item[(i)] the strong and dominant energy conditions hold, and
\item[(ii)] the fall-off condition, $T_i{}^i \to 0$ on approach to $\scri^+$, holds.
\een
If $\scri^+$ is of zero Yamabe class, then either every timelike geodesic in $(M,g)$ is
past incomplete, or else $(M,g)$ is isometric
to the warped product with line element
\begin{equation}\label{eq:split}
ds^2 = - d\tau^2 + e^{2\tau} \tilde h \,,
\end{equation}
where $\tilde h$ a Ricci flat metric on $\scri^+$. In particular, $(M,
g)$ satisfies the vacuum ($T_{ij} = 0$) Einstein equations with cosmological
constant $\Lambda = n(n-1)/2$.
\end{thm}

Here ``maximal" means that $(M,g)$ is not  contained in a larger globally hyperbolic
spacetime.  Note also that the weak energy condition has been replaced by the dominant energy
condition.

\smallskip
\noindent
{\it Comments on the proof:}  One again works in the gauge (\ref{gauge}), where now
$\tilde h_0$ is a metric of zero scalar curvature on $\scri^+$, $\tilde r_0 = 0$.
In this case  (\ref{Hineq})
implies that the inequality $H_t\ge n$  holds to order $t^3$ (see \cite{CG} for an application of this 
in the Riemannian setting).  However, by more refined arguments 
\cite{Anderson, AG} it is possible to show that $H|_{\S_t} \ge n$ for all $t$ sufficiently small.
Theorem \ref{sing2} may then be derived from the following rigid version of Proposition \ref{hawk}.
(The assumption of maximality in Theorem \ref{sing2} is used to show that the local warped product splitting
described below can be made global.)

\begin{prop}\label{rigidity}
Let $M^{n+1}$ be a spacetime satisfying the energy condition
\beq
{\rm Ric}\,(X,X) \ge -n \nonumber
\eeq
for all unit timelike vectors $X$. Suppose $M$ has a smooth
compact Cauchy surface $\S$ with
mean curvature $H$ satisfying $H\ge n$.  
If there exists at least one  past
complete timelike geodesic in $M$,
then there exists a neighborhood of $\S$ in $J^-(\S)$ which is isometric to $(-\epsilon,0]\times
\S$, with warped product
metric, 
$ds^2 = - d\tau^2 + e^{2\tau}h\,,
$ 
where $h$ is the induced metric on $\S$.
\end{prop} 

To prove Proposition \ref{rigidity}, one introduces Gaussian normal coordinates in a neighborhood
$U$ of $\S$ in $J^-(\S)$,
$U = [0,\e) \times \S, \quad   ds^2 = -du^2 + h_u \,.$
 Comparison techniques like those used in the proof of Theorem \ref{sing} imply that each
slice $\S_u = \{u\} \times \S$ has mean curvature $H_u \ge n$.  It can be further shown that
each $\S_u$ 
must be totally umbilic, which leads to the desired warp product splitting.  Indeed if some $\S_u$ were not
totally umbilic, then $\S_u$ could be deformed to a Cauchy surface $\S'$ having mean curvature $H' > n$.
Proposition \ref{hawk} would then imply that every timelike geodesic in $M$ is past incomplete, contrary 
to assumption.  We refer the reader to \cite{AG} for details.\qed   

Thus, Theorems \ref{sing} and \ref{sing2} show that future asymptotically simple and de Sitter
spacetimes obeying appropriate energy conditions are {\it totally} past timelike 
geodesically incomplete,
provided $\scri^+$ is of negative or zero Yamambe class.  The Yamabe class condition may, 
under certain circumstances, be viewed as a topological condition.  Consider for example
the $3+1$ dimensional case, and assume $\scri^+$ is orientable.  Then, by well-known results of 
Gromov and Lawson \cite{GL}, if $\scri^+$ is a $K(\pi,1)$ space, or contains a $K(\pi,1)$ space
in its prime decomposition, then $\scri^+$ cannot carry a metric of positive scalar
curvature, and hence must be of negative or zero Yamabe class.  Hence, in $3+1$ dimensions,
this topological condition may replace the Yamabe class assumption.  

Theorems \ref{sing} and \ref{sing2} say nothing about the occurrence of singularities
in the case $\scri^+$ is of positive Yamabe class.  However the discussion of the previous paragraph
suggests a possible connection between the ``size" of the fundamental group of $\scri^+$
and the occurrence of singularities. 
This view  is supported by the following theorem and subsequent corollary.

\begin{thm}\label{pi1} 
 Let $M^{n+1}$, $n\ge 2$, be a cosmological spacetime obeying  the null energy condition (\ref{ricnull}),
which is asymptotically de Sitter (to both the past and future).
If the Cauchy surfaces of $M$ have infinite fundamental group, then $M$ cannot be 
asymptotically simple, either to the past or the future.  
\end{thm}

The failure of asymptotic simplicity  strongly suggests the development of singularities to both the
past and future: there are null geodesics which do not make it to past and/or future infinity.  The result is well illustrated
by Schwarzschild-de Sitter space, which has Cauchy surface topology $S^1 \times S^2$, as described in section 2.  A further
example of interest having Cauchy surface topology $P^3 \# P^3$  has recently been considered by McInnes~\cite{mcinnes};
this example is obtained as a quotient of  Schwarzschild-de Sitter space.   
The proof of Theorem \ref{pi1} is an application of the
author's null splitting theorem \cite{Gnull}; we refer the reader to \cite{AG,Gcarg} for details.

Recall, if $M$ is future asymptotically simple and de Sitter then the Cauchy surfaces
of $M$ are diffeomorphic to $\scri^+$. Hence, Theorem \ref{pi1}  yields the
following corollary.

\begin{cor}\label{pi1cor}
Let $M$ be a cosmological spacetime obeying the null energy condition (\ref{ricnull}), which is future
asymptotically simple and de Sitter. 
If $\scri^+$ has infinite fundamental group then $M$ cannot be asymptotically de Sitter to the past,
i.e., $M$ does not admit a regular past conformal infinity $\scri^-$, compact or otherwise. 
\end{cor}

To emphasize for a moment the evolutionary viewpoint, consider the time-dual of the above corollary.
Thus, let $M$ be a spacetime obeying the null energy condition which is {\it past} asymptotically
simple and de Sitter, and, for concreteness, take $\scri^-$ to have topology $S^1\times S^2$.
According to work of Friedrich \cite{friedrich}, one can use the conformal Einstein equations to maximally
evolve   well understood initial data on $\scri^-$ to obtain a past asymptotically simple and de Sitter
vacuum solution to the Einstein equations with $\Lambda > 0$, having $S^1\times S^2$ Cauchy surfaces.
In general, one expects singularities to develop to the future. One can imagine black holes formimg, 
with the spacetime approaching something like Schwarzschild-de Sitter space to the far future,
and, hence, admiting a  regular future conformal infinity.  But note that the time-dual of
Corollary~\ref{pi1cor} rules out such a scenario.  Thus, either the singularity that develops must be more
global (leading to a big crunch), or else something more peculiar is happening.  In principle, it is possible
that the spacetime approaches to the far future  something like the Nariai solution \cite{bou}, which is
just a metric product of the $2$-sphere with $2$-dimensional de Sitter space, and hence
is geodesically complete.   Since two spatial
directions in the Narai solution remain bounded toward the future, it does not admit a regular
future conformal infinity; roughly speaking $\scri^+$ degenerates to a one dimensional manifold
in this case.  However, one would not expect such a development to be stable. Thus, it is an interesting
open problem to obtain a detailed understanding (analytic or numerical) of  the generic behavior of the future Cauchy
development in this situation. 

In view of Corollary \ref{pi1cor}, one expects there to exist singularities in the past, in the case $\scri^+$ has infinite
fundamental group.  We conclude with the following theorem, which reinforces this
expectation.

\begin{thm}
Let $M^{n+1}$, $2\le n \le 7$, be a future asymptotically simple and de Sitter spacetime,
with compact orientable future infinity $\scri^+$, which obeys the null energy condition
(\ref{ricnull}).  If $\scri^+$ (which is diffeomorphic to the Cauchy surfaces of $M$) has positive first Betti number,
$b_1(\scri^+) > 0$, then
$M$ is past null geodesically incomplete. 
\end{thm}

Note that if a  Cauchy surface $\S$ contains a wormhole, i.e.,  has topology of the form $N\# (S^1\times S^{n-1})$, then
$b_1(\S)>0$.   The theorem is somewhat reminiscent of previous results  of Gannon \cite{Ga}, which show, in the
asymptotically flat setting, how 
nontrivial spatial topology leads to the occurrence of singularities.  

\medskip
\noindent
{\it Sketch of the proof:} The proof is an application of the Penrose singularity theorem;
see, e.g., \cite[Theorem 1, p. 263]{HE}

Since $M$ is future asymptotically de Sitter
and $\scri^+$ is compact,  one can find in the far future
a smooth compact spacelike Cauchy surface $\S$ for $M$, with second fundamental form which
is {\it positive definite} with respect to the {\it future} pointing normal.  This means that
$\S$ is {\it contracting} in all directions towards the {\it past}.  

By Poincar\'e duality, and
the fact that there is never any co-dimension one torsion, $b_1(\S) >0$ if and only if
$H_{n-1}(\S,\Bbb Z) \ne 0$.  
By well known
results of geometric measure theory (see \cite[p. 51]{lawson} for 
discussion; this is where the dimension assumption
is used),
every nontrivial class in $H_{n-1}(\S,\mathbb Z)$ has a least area
representative
which can be expressed as  a sum of smooth, orientable,
connected, compact,
embedded minimal (mean curvature zero) hypersurfaces in $\S$.  Let $W$
be such a hypersurface; note $W$ is spacelike
and has co-dimension two in
$M$.   As described in \cite{gal:min}, since $W$ is minimal in $\S$, 
and $\S$ is contracting in all directions towards the past in $M$, $W$ must be a past trapped
surface in  $M$.  (Recall \cite{HE, BE, ON} that a past trapped surface is a compact co-dimension  
two spacelike submanifold $W$ of $M$
with the property  that the two congruences of null normal geodesics issuing  to the
past from $W$ have negative divergence along $W$.)  

Since $W$ and $\S$ are orientable, $W$ is two-sided in $\S$. 
Moreover, since $W$ represents a nontrivial element of $H_{n-1}(\S,\mathbb Z)$,
$W$ does not separate $\S$, for otherwise it would bound in $\S$. This implies that
there is a loop in $\S$ with nonzero intersection number with respect to  $W$.
There exists a covering space $\S^*$ of $\S$ in which this loop gets unraveled.
$\S^*$
has a simple
description
in terms of cut-and-paste operations:   By making a cut along
$\Sigma$, we obtain a compact manifold $\S'$
with two boundary
components,
each isometric to $W$.  Taking $\mathbb Z$ copies of $\S'$,
and gluing
these copies end-to-end
we obtain the covering space $\S^*$ of
$\S$.  In this covering, $W$ is covered by $\mathbb Z$ copies of
itself, each one separating $\S^*$; let $W_0$ be one such copy.  
We know  by global hyperbolicity that $M$ is homeomorphic to 
$\Bbb R\times \S$, and hence the fundamental groups of $\S$ and $M$ are isomorphic.
This implies that the covering spaces of $M$ are in one-to-one correspondence
with the covering spaces of $\S$.  In fact, there will exist a covering spacetime
$M^*$ of $M$ in which  $\S^*$ is a Cauchy surface for $M^*$.  Thus, $M^*$ is a 
spacetime obeying the null energy condition, which contains a noncompact Cauchy surface (namely
$\S^*$) and a past trapped surface (namely $W_0$).  By the Penrose singularity
theorem, $M^*$ is past null geodesically incomplete, and hence so is $M$. \qed

\providecommand{\bysame}{\leavevmode\hbox to3em{\hrulefill}\thinspace}

\end{document}